\documentstyle[aps,prl,twocolumn,psfig,draft,floats,amsmath]{revtex}
\begin{document}

\draft
\wideabs{

\title{Diamagnetic Susceptibility and Current Distributions in Granular Superconductors at Percolation}
\author{Henning Arendt Knudsen\footnote{Electronic Address: 
Henning.Arendt.Knudsen@phys.ntnu.no} and Alex Hansen\footnote{Electronic 
Address: Alex.Hansen@phys.ntnu.no}}
\address{Department of Physics, Norwegian University of Science and 
Technology, NTNU, N-7491 Trondheim, Norway}
\date{\today}
\maketitle
\begin{abstract} 
We study a two-dimensional granular superconducting network at the percolation threshold under the influence of an external perpendicular magnetic field. By numerical simulations on the full nonlinear problem, we determine the scaling exponent for the magnetic susceptibility. Further, we report on the scaling properties of the current distribution. The scaling of the current is found to be independent of the value of the magnetic field. Our results are in contradiction with previous numerical results based on linearized equations. We find a value for the susceptibility exponent which does not agree with existing theoretical suggestions, but agrees perfectly with renormalization group calculations.
\\
PACS: 64.60.Ak,75.20.-g,74.50.+r,74.80.Bj
\end{abstract} 
}

\vskip0.5cm
\narrowtext
Superconducting networks have been subject to investigation in various topologies and approximations over the years. The granular superconducting network at the percolation threshold is the main concern of this text. In such a system under the influence of an external magnetic field, the scaling of the magnetic susceptibility obeys a power law. There exist two theoretical predictions for the value of the scaling exponent\cite{DG81,RLT83}. Previously numerical work has been done in order to decide on which theoretical prediction is correct\cite{RH88,HR88}. This work was approximate due to linearizations of the governing equations. We demonstrate in this paper that numerical work on the non-linear equations give different current distributions in the network. There exists further an estimate for the scaling exponent of the susceptibility based on renormalization group calculations\cite{WL91}. Our simulations provide a value for this exponent that differs from the two theoretical suggestions, but agrees with the renormalization group estimate. Attemps to compare with experimental result have been made, but they were not conclusive in favor of any theoretical prediction\cite{MM89a,MM89b}.

Bulk superconductors as well as granular superconducting networks have been modeled by division into discrete nodes between which current can flow. Below a critical temperature $T_c$ the amplitude of the superconducting order parameter is constant over the network, whereas the phase of the order parameter obtains, in principle, a different value at each node. The current flowing between two neighboring nodes is then a function of the phase difference between the nodes.

More generally, disordered or granular superconductors are studied in the same way - as a regular network of nodes. The obvious physical difference in topology between the bulk superconductor and {\it e.g.} the random superconductor insulator-mixture constitutes no mathematical difference. Randomness and disorder are incorporated by allowing each pair of neighboring nodes to be either connected or disconnected. The individual connection between nodes, or the network contacts, are treated as Josephson junctions. Josephson junctions appear in many shapes. In this paper we use the so-called Resistively Shunted Junction(RSJ) model\cite{JLLS84}. Some authors use the term RSJ when referring to the more general Resistively and Capacitively Shunted Junction(RCSJ) model. However, these models are fully equivalent in the physical situation considered here.

In this paper we study the random superconductor-insulator in two dimensions at the percolation threshold\cite{SA94} in a constant perpendicular external magnetic field. In equilibrium there will flow electric current in every closed superconducting loop in the network. This constitutes a critical phenomenon situation where the transport properties become highly nontrivial. Furthermore we emphasis the importance of the nonlinearity of the transport equations. In particular, we find the scaling exponents belonging to the moments of the current distribution, and, more important, the scaling exponent of the diamagnetic susceptibility.

The model is a two-dimensional square array of nodes. Each node is with probability $p=0.5$ connected to each of its four nearest neighbors by bonds. All bonds are modeled as RSJ contacts. The phase of the superconducting order parameter at node $j$ is denoted $\phi_j$. The constant perpendicular magnetic field ${\bf B}$ is represented by the vector potential ${\bf A}$. We choose Landau gauge for convenience. Then the current $i_{jk}$ between two neighboring nodes $j$ and $k$ is
\begin{equation}
\label{eq:current}
  i_{jk} = e_{jk} \sin(\phi_j-\phi_k-A_{jk})
\end{equation}
where
\begin{equation}
  A_{jk} = \int_j^k{{\bf A} d{\bf l}}
\end{equation}
is the line integral from node $j$ to node $k$ of the vector potential. Here $e_{jk}$ is one or zero dependent on whether there is an RSJ contact between nodes $j$ and $k$ or not. The transport equations for the current in the network result from the Kirchhoff current law, for node $j$,
\begin{equation}
\label{eq:kirch}
  \sum_k{e_{jk}\sin(\phi_j-\phi_k-A_{jk})} = 0 .
\end{equation}
The sum is taken over the neighbor nodes. These equations are equivalent to the ones obtained by minimizing the Hamiltonian of the system\cite{A83,HR88}.

The diamagnetic susceptibility is given by\cite{HR88,AM76}
\begin{equation}
\label{eq:suscept}
\begin{split}
  \chi = \frac{\partial M}{\partial B} = & -\frac{1}{S} \sum_{<j,k>}e_{jk}
  \left[ \cos \psi_{jk} \left(\frac{\partial \psi_{jk}}{\partial B} \right)^2  
  \right. \\ & \quad\quad + \left. \sin\psi_{jk} \left(\frac{\partial^2\psi_{jk}}{\partial B^2} \right) \right] ,
\end{split}
\end{equation}
where
\begin{equation}
\label{eq:psidef}
  \psi_{jk} = \phi_j-\phi_k-A_{jk}.
\end{equation}
Here the sum is to be taken over all nearest-neighbor pairs of nodes. $S$ represents the area of the network. The susceptibility is obtained from (\ref{eq:suscept}) in the limit $B\to 0$. In this limit the second term of (\ref{eq:suscept}) vanishes because $\psi_{jk}$ is odd in $B$.

Given a value for the magnetic field, the transport equations (\ref{eq:kirch}) can be solved for $\phi_j$, the phase of the order parameter at each node $j$. In turn the current in each bond is found from (\ref{eq:current}). The current distribution thus found is not multifractal. This is surprising as previous work on the linearized case showed multifractality\cite{M83,F88,SA94}. The concept of multifractals can be explained as follows. Having calculated the currents in the network bonds, the moments $M_l$ of the current distribution are found,
\begin{equation}
\label{eq:moment1}
  M_l = \sum_{<j,k>}{i_{jk}^l}
      = \sum_{<j,k>}{e_{jk}[\sin(\phi_j-\phi_k-A_{jk})]^l}.
\end{equation}
The sum is over all nearest neighbor pairs of nodes. Now, the question arises about how these moments scale with the linear size $L$ of the networks considered. If each moment obey a power law:
\begin{equation}
\label{eq:moment2}
  M_l \sim L^{\tau_l},
\end{equation}
where the exponents $\tau_l$ do \emph{not} have the form $\tau_l=al+b$, then the current distribution is multifractal.

Hansen and Roux\cite{HR88} have calculated the multifractal exponents resulting from the linearized version of (\ref{eq:kirch}),
\begin{equation}
\label{eq:linkirch}
  \sum_k{e_{jk}(\phi_j-\phi_k-A_{jk})} = 0 .
\end{equation}
It would seem that this approximation is valid for small $B$. We are interested in the behavior of the system in the limit $L\to\infty$. This linearization process is in fact so that we first take the limit $B\to 0$ and then the limit $L\to\infty$. That is \emph{not} the same as taking $L\to\infty$, then $B\to 0$. We report the results from the solution of the non-linear equations (\ref{eq:kirch}) and find a different set of exponents valid for all $B$. Likewise in the linear case, (\ref{eq:linkirch}), a similar linearization of (\ref{eq:suscept}) and (\ref{eq:moment1}) shows that the diamagnetic susceptibility is equal to the second moment of the current distribution. Whence the susceptibility scales as does the second current moment.

This easy relationship between the current distribution and the susceptibility is a direct result of the linearization. There is no reason why this should hold also in the non-linearized case. Actually another set of coupled equations must be solved for $({\partial \psi_{jk}}/{\partial B})$ which appears in (\ref{eq:suscept}), before the susceptibility can be calculated. This set of equations is found by realizing that the left hand side of (\ref{eq:kirch}) is in fact an invariant with respect to the magnetic field $B$. Thus the derivative of this expression with respect to $B$ must be zero, for node $j$ having $k$ neighbors
\begin{equation}
\label{eq:phider}
  \sum_k{e_{jk}\cos\psi_{jk}\frac{\partial \psi_{jk}}{\partial B}} =0.
\end{equation}
By the solution of (\ref{eq:kirch}) the cosine factor in (\ref{eq:phider}) is evaluated, and the equations become linear in the unknown partial derivatives.

Numerically we have analyzed ensembles of seven different system sizes. The networks in the ensembles are at the percolation threshold, meaning that each bond connecting a neighboring pair of nodes is drawn by chance to be present with probability $p=p_c=0.5$. The equations are solved using the conjugate gradient iterative scheme. The number of networks in the ensembles are as follows; $L=16: 20000$, $L=25: 4000$, $L=40: 2000$, $L=64: 800$, $L=100: 130$, $L=140: 60$, and $L=200: 20$. All the networks were analyzed for different values of the magnetic vector potential $A$. We show the calculated current moments in figure \ref{fig1}. Each of the curves correspond to one network size. It is intuitively clear that a larger network will carry more current than a smaller. On the other hand, that the exact scaling relation of $M_1$ as a function of $L$ should be a power law is not so clear. We also do not have much intuition for the value of the scaling exponent. The data in figure \ref{fig1} can be rescaled with respect to their maximum value, shown in figure \ref{fig2}.

The results show an excellent example of data collapse. This tells us that the scaling relation of the first current moment is the same for all values of the vector potential. The scaling exponent from this scaling is found from figure \ref{fig3}. The upper curve corresponds to the scaling of the first current moment. The power law fit is very good, and the exponent is $\tau_1 = 2.01$. The middle and lower curve in figure \ref{fig3} correspond to the scaling of the second and fourth current moment respectively. Our work show that the scaling of all the current moments are independent of the value of the magnetic field. The scaling exponents found are summarized in table \ref{table1}. Here we also quote previous results based on the linearized equations\cite{HR88}.

\begin{table}
\caption{The table gives the scaling exponents of the moments of the current distribution. The results are given for both previous linear results and present nonlinear results.}
\label{table1}
\begin{tabular}{rll}
  Moment $l$ & $\tau_l$ (linear) & $\tau_l$ (non-linear) \\
\hline
  1          & 2.26              & 2.01            \\
  2          & 3.03              & 2.00            \\
  4          & 4.75              & 2.00            \\
  6          & 6.68              & 2.01            \\
  8          & 8.61              & 2.01            \\
\end{tabular}
\end{table}

These two sets of exponents differ drastically. The linearized equations give rise to a multifractal current distribution. Our calculations on the non-linear equations show that those results must be wrong. The current distribution is not multifractal. It is not even fractal as the exponents within reasonable error limits are equal to the integer number two.

The magnetic susceptibility $\chi$ of the networks is shown in figure \ref{fig4} as a function of the linear size $L$. A best power law fit gives the scaling exponent $b/\nu=0.90\pm0.03$ where the error bar comes from fitting the data. In the same figure we show with a dashed line the previously established result from the linearized numerical work, $b/\nu=1.03$\cite{RH88,HR88}.

The previous result is in perfect agreement with the expression attributed to de Gennes\cite{DG81},
\begin{equation}
\label{eq:DG}
  b/\nu=2-t/\nu=1.025 .
\end{equation}
Rammal \emph{et al} have given another suggestion\cite{RLT83},
\begin{equation}
\label{eq:RLT}
  {b}/{\nu}=2-{t}/{\nu}+{\beta}/{\nu}=1.129 .
\end{equation}
Here $t=1.300$, $\nu=4/3$, and $\beta=5/36$ are exponents known from percolation theory\cite{SA94}; conductivity critical exponent, correlation length critical exponent, and the critical exponent governing the vanishing of the density of the infinite cluster respectively. Wang and Lubensky has later calculated $\chi$ using a cumulant expansion method\cite{WL91}. This renormalization group approach gives the value $b/\nu=0.91\pm0.03$, which is in perfect agreement with the present simulation. The experimentally quoted values are 0.99 and 1.09 \cite{MM89a,MM89b}, which are values comparable to all of the theoretical and numerical numbers. However, it seems not to be possible to draw conclusions on this basis.

Our results are interesting for two major reasons. One being a demonstration of the dangers of linearizing a real nonlinear problem. This is equivalent to an interchange of limits as discussed above. Our results indicate that this interchange is not allowed, and the penalty is wrong answers. The other being that the value for the susceptibility critical exponent found is in contradiction with both existing theoretical predictions. The previous verdict was in favor of (\ref{eq:DG}), but now the case must be reopened. It might just well be the case that this exponent describes a fundamentally different property of the percolating system, other than the well known exponents $\nu$, $t$ and $\beta$. Whence one should not expect it to be expressible in turns of them. Generally it is known that renormalization group calculations can give reasonably precise estimates for critical exponents even though short cumulant expansions are used. At the same time standard series expansion methods of governing equations are known to fail in critical phenomenon situations. Therefore the result of Wang and Lubensky seems more plausible, and the fact that numerical simulations gives the same result strongly indicates that this is indeed the correct value for the exponent.

In summary, we have studied the granular superconducting network at the percolation threshold under the influence of an external perpendicular magnetic field. We find the scaling exponent for the magnetic susceptibility to be $b/\nu=0.90$ which differs both from existing theoretical suggestions and previous numerical work, but agrees with renormalization group calculations. Further, the moments of the current distribution is found to scale with exponent two, \emph{i.e.}, with the area of the network. This scaling is independent of the value of the magnetic field. This contradicts the previous studies of the linearized case which gives a multifractal current distribution. We conclude that this linearization is not legal and is equivalent to interchange of
the limits $B\to0$ and $L\to\infty$.


\begin{figure}
\caption{The first current moment as a function of the magnetic vector potential $A_0$. We show the moments for different network sizes $L$ as shown in the legend of figure \ref{fig2}. 
\label{fig1}
}
\end{figure}
\begin{figure}
\caption{The same data as in figure \ref{fig1}, where each curve has been normalized with respect to its maximum value. We observe the good quality of the data-collapse.
\label{fig2}
}
\end{figure}
\begin{figure}
\caption{Scaling of first, second, and fourth moments of the current distribution. The straight lines are best power law fits. The scaling exponents are found in table \ref{table1}.
\label{fig3}
}
\end{figure}
\begin{figure}
\caption{The diamagnetic susceptibility $\chi$ is shown as a function of the network size $L$.
\label{fig4}
}
\end{figure}
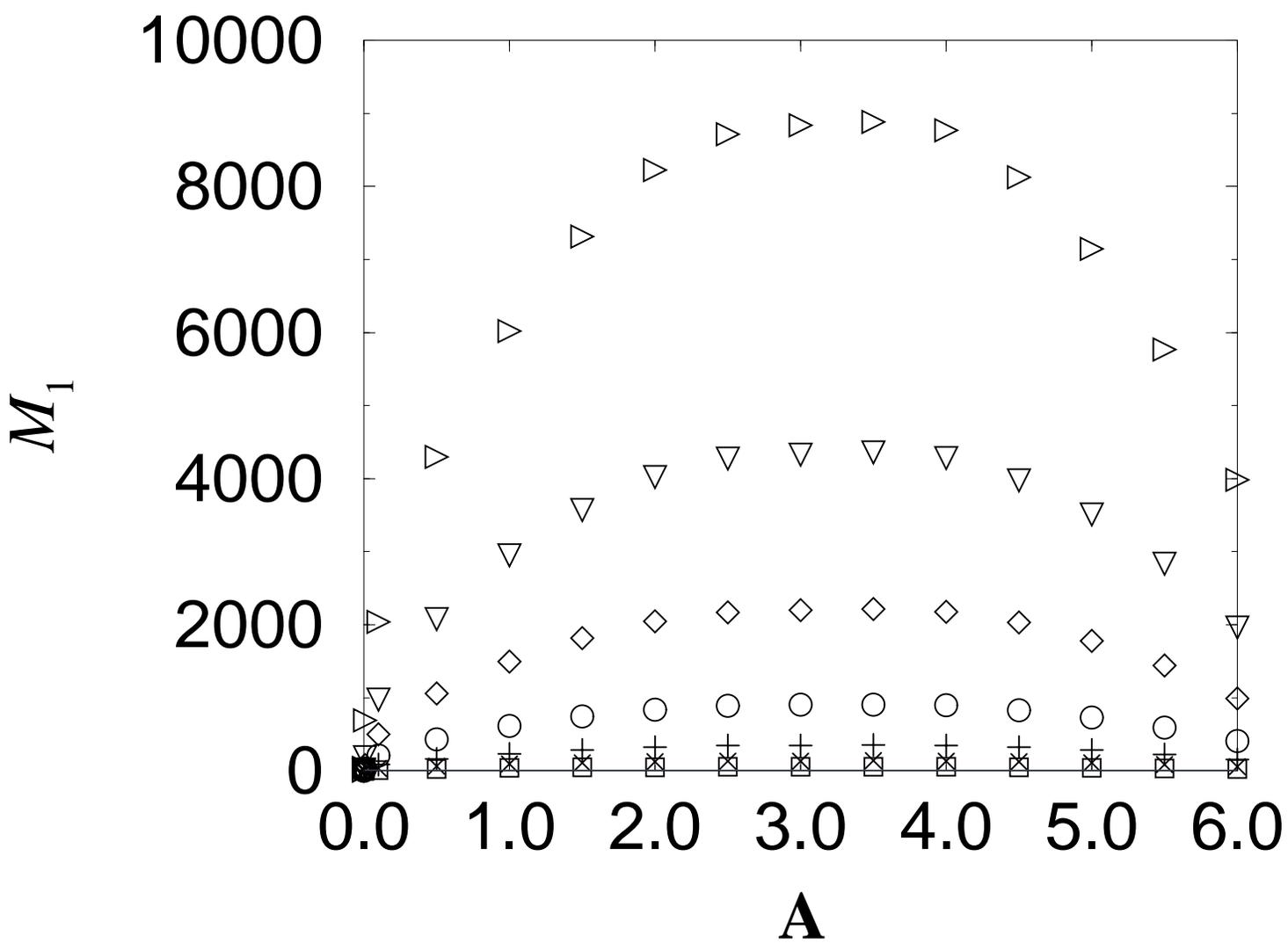
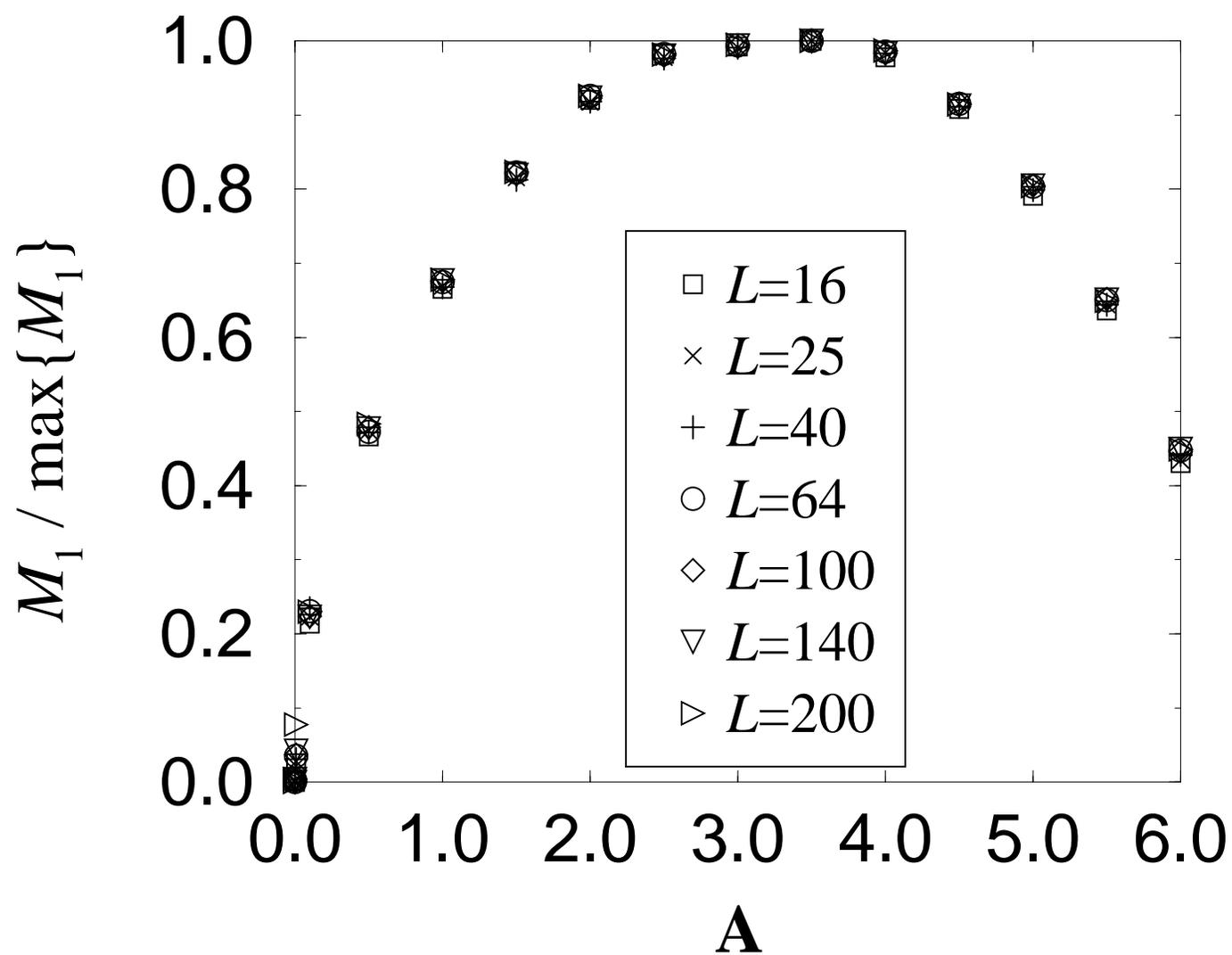
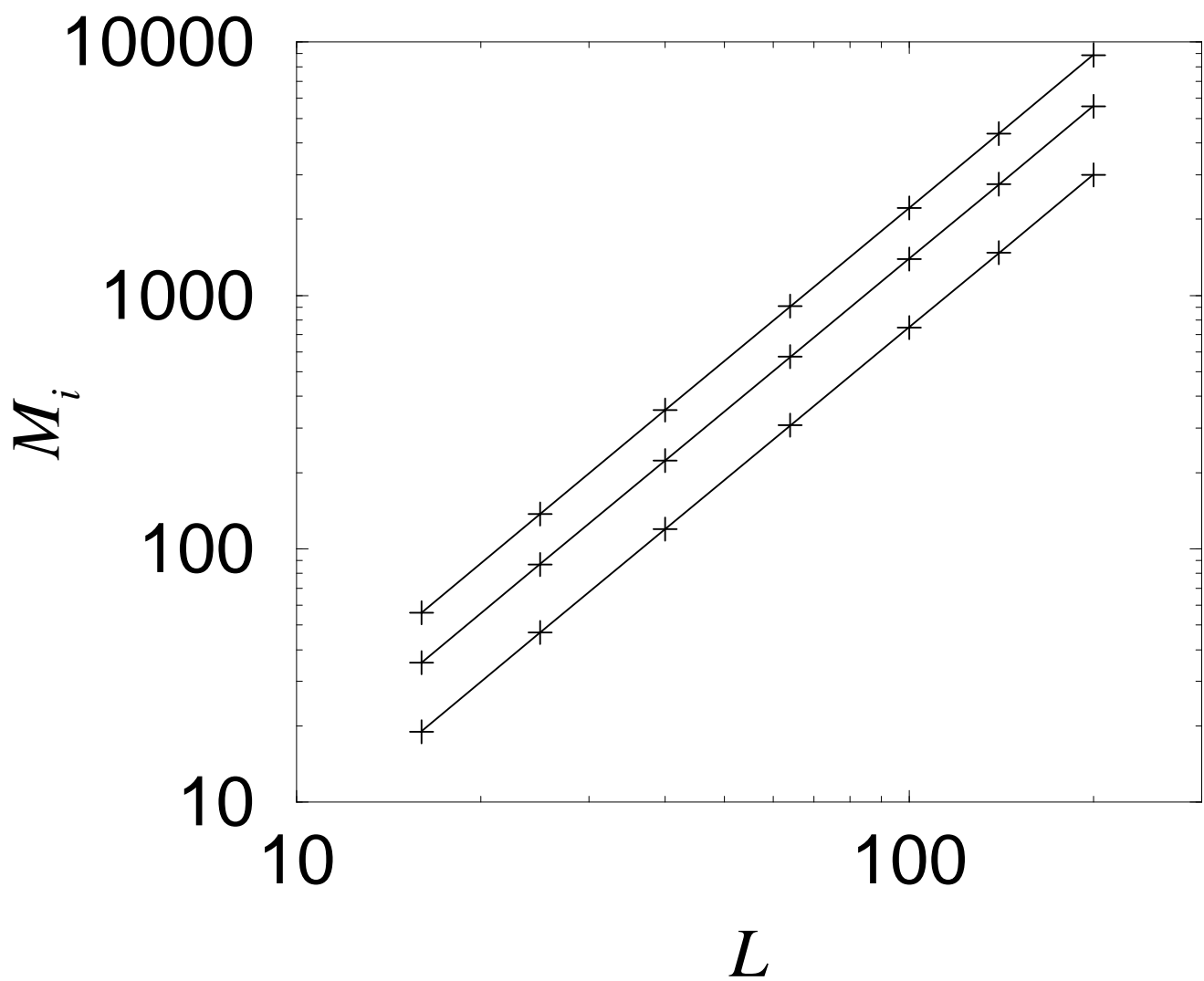
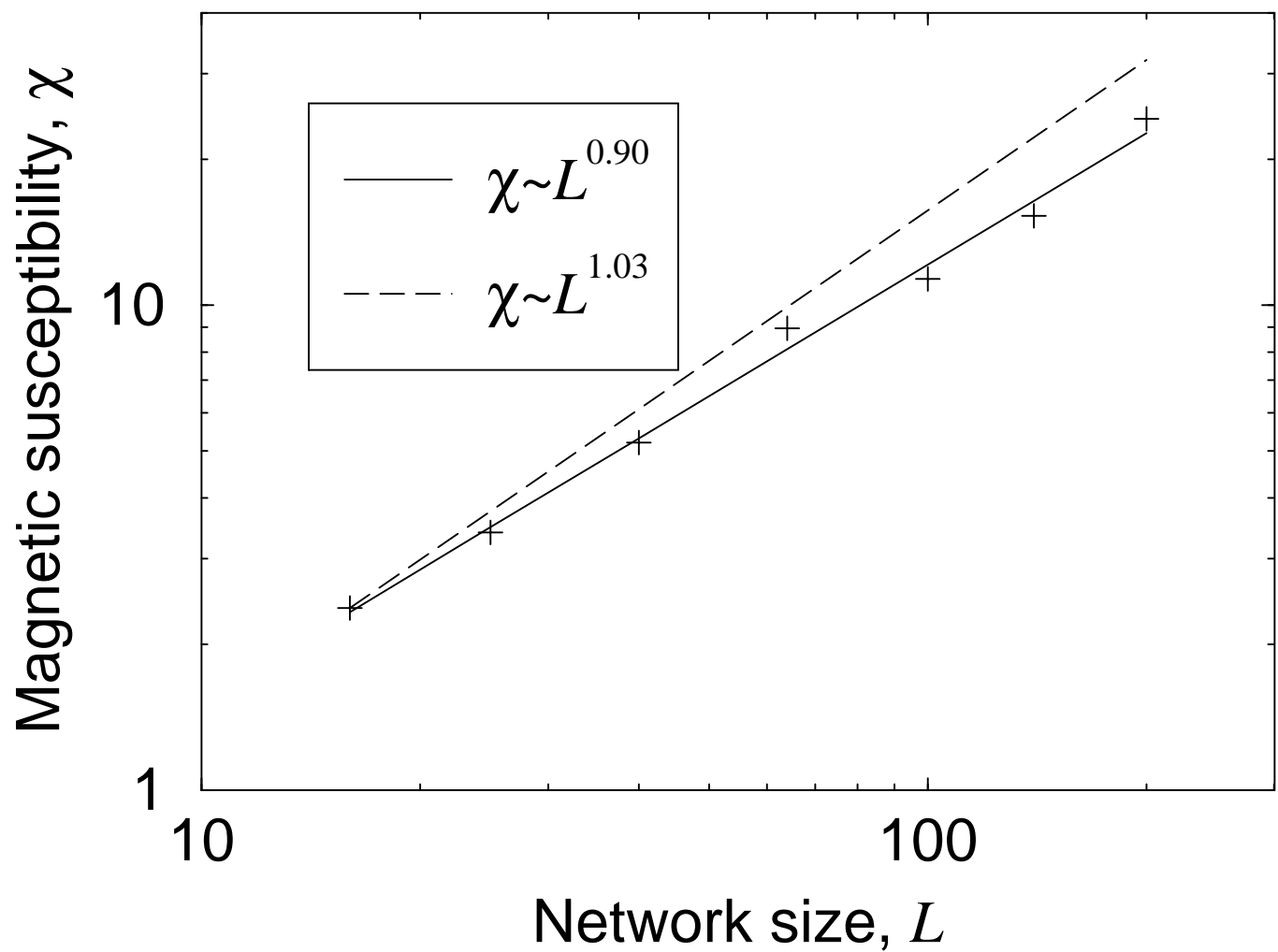
\end{document}